# Three-dimensional Orientation Sensors by Defocused Imaging of Gold Nanorods through an Ordinary Wide-Field Microscope

Tao Li, Qiang Li, Xiao-Jun Chen, Yi Xu, Qiaofeng Dai, Hai-Ying Liu, Sheng Lan and Li-Jun Wu*

*Laboratory of Photonic Information Technology, School for Information and Optoelectronic Science and Engineering, South China Normal University, Guangzhou 510006, P.R. China*

**Corresponding Author**

*Email: ljwu@scnu.edu.cn

Address:

High Education Mega Centre 378 Waihuan West Road,

Guangzhou 510006, P.R. China

Phone: 0086-020-39310366

Fax: 0086-020-39310083

**Abstract:**

Gold (Au) nanoparticles particularly nanorods are actively exploited as imaging probes because of their special nonblinking and nonbleaching absorption, scattering and emitting properties that arise from the excitation of surface plasmons. Herein, we report a novel orientation sensor at nanoscale by defocused imaging of single Au nanorods (AuNRs) through an ordinary wide-field optical microscope. By simultaneously recording defocused images and two-photon luminescence intensities for a large number of single AuNRs, we correlate their defocused images with their three-dimensional spatial orientations. As many AuNRs can be monitored in parallel, defocused imaging of single AuNRs is a high throughput sensing technique that allows us to obtain their spatial orientations within one frame in situ and real-time. The probe size can be down to several nanometers, which is highly desirable in order to minimize any potential interference from the probe itself. Furthermore, the sensing property is insensitive to the excitation polarization and the distribution of the probe aspect ratio, which allows AuNRs of any length in a proper regime to be used as orientation sensors without changing the laser frequency and polarization. These unique features make the orientation probes proposed here outstanding candidates for optical imaging and sensing in material science and biology.

Optical probes for nanoscale orientation sensing have attracted much attention in the field of single-molecule spectroscopy [1,2]. On one hand, local orientation sensors are very important in material research, e.g., liquid crystal orientation [3], the dynamics of polymers near the glass transition temperature [4] and the conformation and rotation of single polymer chains [5, 6]. On the other hand, in biology research, the spatial orientation of the probes is of great significant to understand certain vital biological mechanisms, such as the structural dynamics of myosin V [7], the structure of Dendra-2-actin in fibroblast cells [8], the rotational motions of nano-objects in live cells [9], or the molecular dynamics of early virus-host couplings for cell infection [10], etc. Most of the currently employed probes are fluorescent labels, including fluorescent polymers [5], organic dyes [7,9], and inorganic semiconductor nanoparticles [10], etc. However, time-dependent fluorescence fluctuations (photoblinking) and a limited measurement time due to irreversible photochemical changes (photobleaching) of these probes seriously restrict their applications for imaging [7,11,12]. Although core-shell structured semiconductor nanoparticles could alleviate photobleaching in some degree [13], the toxicity of these nanoparticles could be a fatal drawback for their practical applications in life science [14].

Noble metal nanoparticles (NPs) exhibit extraordinary plasmonic properties, great photostability, excellent biocompatibility and nontoxicity, and thereby could be an alternative label to organic dyes or quantum dots [15,16]. In addition to generating magnitudes larger absorption and scattering cross-sections than that of dye molecules [17], the collective oscillation of their localized surface plasmons (LSPs) also gives rise to a strong polarization of the fluorescence, the absorbed and scattered light because of their inherent anisotropy [2, 18-20]. This polarization characteristic can be tuned by the shape of the NPs which allows for orientation imaging [21]. One type of the most interesting metallic NPs with an anisotropic shape is Au nanorods (AuNRs), in which the longitudinal and transverse localized surface plasmon resonant (LSPR) modes are parallel and perpendicular to the rod length direction respectively. For these reasons, Link et al. has proposed an orientation sensor based on the plasmonic absorbance of AuNRs [1]. Utilizing this kind of photothermal imaging technique, however, is difficult to extract the three-dimensional (3D) orientation information of AuNRs within one frame. It requires a combination of a time-modulated heating beam and a probe beam which complicates the measuring process. Furthermore, the probe AuNR is easy to be melted and reshaped as it is based on absorption [22].

In the conventional imaging method, the 3D orientation of the specimen cannot be resolved within one frame because the scattered or emitted light from all the dipoles on the focal plane are focused into a single Airy disk. If an aberration (slight shift of the dipole away from the focal plane) is deliberately introduced to the imaging system, direct observation of the spatial distribution of the emitted or scattered field (defocused image) of single dipoles becomes available. This technique is based on the electron transition dipole approximation and the fact that the dipole radiation exhibits an angular anisotropy. Comparing with the conventional orientational imaging with two crossed polarizers, the defocused imaging method allows direct access to the 3D orientation of the emitters or scatters as well as their radiation characteristics within one frame [23, 24]. Very recently, based on the strong scattering of SPs, Yeung et al resolved 3D orientations of single AuNRs by using a standard optical dark-field microscope through deciphering the defocused darkfields images [25]. Nonetheless, this scattering-based techniques is complicated by the fact that many other biological objects also scatter strongly giving rise to a large background and hence a decrease in sensitivity. In addition, the scattering cross-section scales with the radius ($r$) of the NP according to $r^6$ [17]. Metallic NPs much smaller than 50 nm in diameter are therefore not detectable by this kind of dark-field single-particle imaging technique [17]. This prohibits a reduction in probe size.

On the other hand, when the radiation emitted by a single fluorophore can be approximated as a dipole, it is very convenient to utilize defocused imaging to investigate its emission characteristics [26]. Fluorescence from metal materials has been observed for more than thirty years but has not attracted much attention due to very low quantum yields of ~$10^{-10}$ [27]. Recently, fluorescence with much higher quantum yield (>$10^{-5}$) was observed both in metal nanoclusters (NCs) (< 2 nm) [28] and in small metal NPs (<100 nm) [29-31] The photoluminescence (PL) of Au NCs could be assigned to the radiative recombination of Fermi level electrons and *sp*- or *d*-band holes created after photoexcitation [32,33]. Although the size of the Au NC is as small as less than 2 nm, which is highly desirable for biosensing, unfortunately, at the single-molecule level, its fluorescence suffers from photoblinking and photobleaching [34]. When the size of the NP is increased to be able to support a "plasmon" characteristic of a large number of free electrons, its fluorescence could be assigned to radiative emission of excited LSPR, in which the exited *d*-band holes would recombine nonradiatively with *sp*-electrons and emit particle plasmons [30]. When the aspect ratio of AuNR is less than 5, its excited plasmon mode is fundamental and exhibits dipolar character [35]. Thus its emission can be approximated as a dipole because of the collective oscillation of free electrons.

As the response increases quadratically with intensity in two-photon processes, the two-photon luminescence (TPL) provides greatly enhanced sensitivity for investigation of individual AuNRs and imaging of their TPL is expected to be a promising tool for optical probing [35]. In fact, TPL is one of the most-extensively studied processes related with the application of AuNRs [36, 37]. Despite these advantages, it cannot be applied in the defocused imaging technique because of the tight-focus exciting requirement. On the other hand, AuNRs can also be excited by single-photon process with a proper wavelength [29, 38]. In this paper, we take the advantages of single-photon luminescence (SPL) defocused images to validate AuNRs as new probes for 3D orientation sensing. By simultaneously recording defocused images and TPL intensities upon tuning the incident polarization for a large number of single AuNRs with aspect ratios of 1.2-4, we correlate their defocused images with their 3D spatial orientations. This allows us to unambiguously derive the spatial orientation of single AuNRs and their radiation characteristics within one frame. The technique we propose here is highly reliable and offers great improvements in comparison to previously reported methods, including fast full 3D angle resolving capability, freedom from photobleaching and photoblinking, high throughput for probing, small probe size, and flexible excitation. These unique features make it an outstanding candidate for optical imaging and sensing in material science and biology.

AuNRs were prepared chemically in aqueous solutions by a seed-mediated method [39]. Briefly, a seed sphere solution with 3-5 nm Au nanocrytals was generated by adding ice-cold $NaBH_4$ to a solution of $HAuCl_4$ in the presence of cetyltrimethylammonium bromide (CTAB). This light-brown seed solution was vigorously stirred for 30 mins and then kept at room temperature before further used. To make a growth solution, 400 μL of 25 mM $HAuCl_4$ and 10 mL of 0.2 M CTAB were mixed with 10 mL of deionized water (18 MΩ), followed by the addition of 100 μL of 4.0 mM $AgNO_3$ solution. After then, 200 μL of 0.08 M ascorbic acid was added and gently stirred for 30s, which changed the color of the growth solution from orange to transparent. The growth of AuNRs was initiated by injecting 24 μL of the seed solution into the growth solution. After leaving for 24 hrs at room temperature, the AuNRs could be purified from the excess surfactant solution by centrifugation (13000 rpm for 15 min). The aspect ratio of AuNRs can be controlled by tuning the concentration of $AgNO_3$ in the growth solution.

Figure 1 (a) illustrates the normalized UV-Vis absorption spectra of the fabricated AuNRs with three typical aspect ratios. The transverse plasmon resonance band appears to be positioned at approximately 515 nm, while the maximum of the longitudinal plasmon resonance wavelength red shifts with increasing the aspect ratio. Figure 1 (b)–(d) show the corresponding transmission electron microscopy (TEM, JEM-2100HR) images. The aspect ratio was estimated to be 2.1, 2.5 and 3.5 respectively. The AuNRs were deposited on 0.2 mm thick cleaned glass coverslips at a low concentration (the coverage is less than one rod per $\mu m^2$). They are well-separated from each other.

The details for our experimental setup to investigate the defocused images of single AuNRs are similar to what shown in Ref [40, 41]. In brief, the samples were excited by a diode-pumped solid-state laser (532nm, 400W/cm$^2$, Coherent) that is directed through a Zeiss 100×/1.3NA oil immersion objective. A half-wave plate was inserted into the incident circuit to rotate the polarization of the excitation with respect to the laboratory reference system from 0° to 180° with 10° increments. In order to improve the signal-to-noise ratio, the size of the beam from the laser was reduced before it entering into the microscope. With a long pass filter to block the excitation light in the detection path, the PL from AuNRs was collected by the same objective and detected using an intensified charge-coupled device (CCD) (Carl Zeiss) camera, as shown in Supporting Information (Figure S1). The defocused image was taken by moving the objective around 1 μm toward the sample after a clear diffraction-limited fluorescence image appeared. All the fluorescence images were recorded at room temperature. The exposure time for each image was set to be 200 ms in order to obtain enough signal-to-noise ratio. For TPL polarization dependence measurement, a 800 nm laser pulse (130 fs and 76 MHz, 10 nJ/mm$^2$) from a Ti: sapphire oscillator (Mira 900, Coherent) was used as the excitation source. The excitation polarization was controlled by a half-wave plate operating at 800 nm in the incident circuit. The SPL (TPL) spectrum of single AuNRs in focus was measured by an Andor electron-multiplying CCD (EMCCD, DU970N-BV and DU-897E) together with a 540-910 nm (370-720 nm) bandpass filter in the output circuit. The details for the measurement are described in the caption of Figure S1. We normally choose the target AuNR by tracking its TPL intensity upon rotating the excitation polarization at first. After then, we change the excitation wavelength to be 532 nm and record the defocused images. The optical circuit has to be adjusted very carefully to ensure that we are probing the same single AuNR during these two processes.

The program used to simulate the defocused images is based on the multi-dimensional (3D here) dipole model developed by Enderlein etc [26, 42]. The definition of the angular coordinates is shown in Figure S2, where Y is the out-of-plane angle (inclination angle between the rod length direction and optical axis, and b the in-plane angle.

It has been broadly accepted that the TPL intensity is maximized when the polarization of the excitation is parallel to the long axis of AuNRs [36, 37] as the excitation of interband absorption and the longitudinal surface plasmon of AuNRs can result in enhanced local electric fields which amplify both exciting and emission fields. It follows a $\cos^4\theta$ dependence (θ is the angle between the excitation polarization and the long axis of the rod) [35]. This polarization characteristic of the TPL intensity can be applied to determining the orientation of the NR. Here we utilize this property to verify the orientation of the AuNRs derived from the defocused imaging technique. This process is not as complicated as fabricating an identification pattern on the glass substrates to position the single AuNRs [1, 43]. We start with a simple case in which the long axis of the AuNRs is nearly parallel to the cover slide.

Typical TPL spectra for two single AuNRs with aspect ratios 2.0 and 3.5 are shown in Figure S3 (a).

To avoid the influence from a second harmonic generation of AuNRs, the TPL intensity was obtained by integrating over the wavelength range from 570- to 640 nm. Figure 2 (a) plots its traces for a representative single AuNR upon adjusting the excitation polarization, with the measured and simulated defocused images displayed in the inset. Unless specifically stated otherwise, the black and white defocused images correspond to the measured results and colorful ones to simulated results. As seen, the intensity traces for TPL can be fitted to be a $\cos^4(\theta-\beta)$ function. The in-plane orientation of the AuNR can thus be deduced to be 113.6°. The corresponding defocused image shows a bilaterally symmetric two-lobe pattern. By fitting the experimental parameters into the simulation program, the angle $\beta$ is derived to be 113°. The two closely consistent results indicate that the defocused image of single AuNRs can be applied as a novel orientation probe.

To back our conclusions with a statistically meaningful dataset, we investigated the polarization characteristic of TPL for 30 single AuNRs and simultaneously recorded their defocused images. Figure 2 (b) illustrates the correlation of orientation angles for 30 AuNRs determined independently by these two techniques. The largest deviation is 14°. Several factors, such as the 10° step to rotate the excitation polarization, the difference between the optical circuits for single-photon excitation (532 nm) and two-photon one (800 nm) and the process of fitting defocused images could introduce the deviation. However, the excellent correlation between the orientation angles obtained by the two techniques confirms that the spatial orientation of single AuNRs can be unambiguously determined by the defocused imaging technique.

When the AuNR is orientated out of plane, i.e., 0°<Ψ<90°, we can also decipher its spatial orientation by fitting the defocused images. A map of simulated defocused images from single AuNRs with various 3D orientations is shown in Figure S4. Obviously, each specific group of angles is correlated with a unique defocused image. Therefore, the random orientation of AuNRs could be easily determined by referring to their corresponding emitting field map. It is worth noting that, for the out-of-plane orientated AuNRs, the most two obvious characteristics of the defocused images are the two-fold symmetry breaking of the brightness of the high-order diffraction ring (when 90°>Ψ>30°) and the bright spot at the center deviating away from the central position (when 0°<Ψ<30°). As is known, the 2D in-plane orientation (the projection of the oscillation dipole onto the image plane) of AuNRs could be accurately determined via modulating the excitation polarization [25]. For the out-of-plane angle, we utilize a parameter $DOP=(I_{max}-I_{min})/(I_{max}+I_{min})$, where $I_{max}$ and $I_{min}$ are the maximum and minimum luminescence intensities at two perpendicular polarization directions, to qualitatively analyze. When the AuNR is orientated out of plane, its *DOP* should be smaller than that for the in-plane positioned AuNR. Column (a) in Figure 3 shows the measured and simulated defocused images of three typically selected AuNRs. Through fitting the field distribution pattern, the 3D angles of the AuNRs are measured to have polar angles of 0°, 25°, 81° and azimuthal angles of 96°, 26°, -72°. The azimuthal angles, i.e., in-plane projection angles 96° and 26° perfectly match those obtained through measuring the excitation polarization dependence of TPL as demonstrated in column (d) of Figure 3.

Furthermore, as evidently shown in Figure 4, the *DOP* of the TPL from AuNR2 is smaller than that from AuNR1, revealing that AuNR2 is really orientated out of plane. The defocused image of AuNR3 is nearly isotropic and its TPL is independent on the excitation polarization. There are two possibilities for the appearance of this defocused pattern. One is that the long axis of the AuNR is almost perpendicular to the glass substrate. It is also possible that AuNR3 is an aggregation of several AuNRs.

These perfectly matched results reveal that from the defocused imaging analysis not only the accurate 2D in-plane projection information can be deduced, the tilt angle information can also be resolved directly, indicating that this technique is very informative. More importantly, it can be utilized to obtain live images and simultaneously track and monitor multiple dynamic events with fast data acquisition rate. As large numbers of AuNRs can be monitored in parallel within one frame, it is capable of achieving high throughput for sensing. Therefore, the orientation sensor based on defocused imaging of single AuNRs is an ideal candidate for optical imaging and sensing various reaction processes in material science and biology.

In the following, we investigate the influence of the excitation to examine the flexibility of this technique. The excitation polarization plays an important role to excite the TPL of AuNRs. The intensities integrated over the whole defocused images and the wavelength range from 540- to 910 nm are plotted as a function of the incident polarization angle in Figure 5 (a) and (b) for two randomly selected AuNRs. The insets show the defocused images of the single AuNRs for different polarization angles. It can be clearly seen that although the intensity response of the defocused images is accordingly varied with the incident polarization, the pattern of the defocused images, however, remains unchanged regardless of the NR orientation. This result was unexpected. When the excitation is polarized along the short axis of AuNRs, we thought only the transverse mode could be excited as the excitation 532 nm is very close to its SPR. Thus we expected to see a 90$^o$ rotation of the defocused pattern upon tuning the incident polarization from parallel to perpendicular to the long axis of AuNRs. It seems there are some other mechanisms behind our experimental results.

As we know, the increased efficiency of the fluorescence from AuNRs can be contributed either to the enhancement of the incoming exciting light (absorption) or the outgoing emitted light (emission) via the excitation of SPR [29]. The PL from Au thin film or NPs has been assigned to radiation recombination of excited electrons in the *sp*-band and holes in the *d*-band [32, 33]. The optical transitions preferentially occur near the *X* and *L* symmetry points of the Brillouin zone [32, 36, 37]. As the $\Gamma L$ direction is orientated between the long and short axes of AuNRs grown from the seed-mediated method [36, 37, 39], the emission from the $\Gamma L$ direction can still be excited when the incident polarization is along the short axis of the AuNR. As shown in Figure S3 (b), the SPL spectrum exhibits a broadband emission from 560- to 660 nm, which is close to the SPR of the longitudinal mode. The emission is thus possible to be enhanced by the longitudinal SPR. On the other hand, the enhancement of the absorption seems playing a minor role here. Therefore, the emission from the dipole oscillating along the long axis seems always dominant and results in changeless defocused patterns upon tuning the incident polarization. The detailed mechanism is still under investigation. However, this phenomenon means that any polarization can validate the probe within one frame and it is not necessary to insert a half-wave plate into the incident circuit to optimize the polarization direction. Furthermore, unlike scattering, the PL from AuNRs is not critically dependent on the excitation wavelength as the absorption band is relatively broad [44]. The pattern of the defocused image is thus expected to be insensitive to the excitation.

When the aspect ratio (*R*) of the AuNR is less than 5, the fundamental plasmon mode is dipolar in character, and the resonance peak of the mode shifts to the longer wavelength as the aspect ratio increases [45]. When $R>5$, multipolar plasmon modes become dominant and multipolar oscillations would reduce the spatial confinement and produce a reduction of the field enhancement. As the defocused imaging method is based on the dipole approximation model, we controlled the aspect ratio of AuNR to be less than 5, in which the emission can be approximated to be from three perpendicularly-positioned single dipoles. Figure 6 (a) displays the defocused images from six single AuNRs with aspect ratios of 1.2, 2.0, 2.5, 3.0, 3.5, 4. The corresponding orientation information obtained from the defocused images and polarization characteristics of their TPL are shown in Figure 6 (b). Obviously, they are in good agreement.

There is normally a distribution of aspect ratios in the fabrication of AuNRs even when the growth parameters have been controlled very strictly [46]. As we know, the longitudinal SP resonance of the AuNRs is significantly dependent on their aspect ratios. For previously proposed orientation sensing techniques based on such as photothermal imaging, the excitations have to be tuned when the aspect ratio of the probes varies [1, 35]. However, the SPL emission spectrum is quite broad as shown in Figure S3 (b). The enhancement for the outgoing emitted light is expected to be valid for a large range of $R$. Furthermore, the brightness symmetry of the defocused image is very easy to be broken if there is a little difference between the emission from the two perpendicularly orientated dipoles. Therefore, in a large range of aspect ratios, the defocused image of AuNR exhibits anisotropy and possesses orientation sensing function. The fabrication tolerance for this probe is thus very large. More importantly, we can apply a same light wavelength to excite the probe, which is highly desirable for optical imaging and sensing.

As smaller particles are more attractive in the language of minimizing potential interference from the probe itself, we exploit the size effect on the defocused images of AuNRs. We fabricated much smaller AuNRs according to the method proposed in reference [47]. Figure 7 (a) exhibits a representative TEM image. The diameter and length of the AuNRs was determined to be about 9 nm and 20 nm in average, which corresponds to an effective diameter of 12 nm for a sphere. A typical defocused image taken from this sample is demonstrated at the top of Figure 7 (a). Evidently, it exhibits similar pattern as those from larger AuNRs shown in the above context. We also tracked the TPL intensity upon tuning the excitation polarization, as displayed in Figure 7 (b). It was found that the orientations obtained by these two techniques are consistent, revealing that defocused imaging can measure the spatial orientation of single AuNRs as small as ten nanometers.

At last, we discuss the influence of the defocusing distance as it affects the defocused pattern seriously [25]. Figure 8 shows a series of seven defocused images for the same AuNR when the defocused distance was varied from 0 to 1.2 μm. The upper panel shows the experimental results and the lower panel the simulated images. We can find a fair correspondence between them. Obviously, when the defocused distance is in the range of 0.5-1.2 μm, the defocused patterns exhibit anisotropy and the in-plane orientation can be derived easily. Figure S5 displays more detailed simulated results, in which all the in-plane orientations of single AuNRs can be deduced when the defocused distance is in the range of 0.5-1.2 μm. However, only the brightness symmetry of the high-order diffraction ring and the bright spot near the center of the defocused image can provide accurate information of the out-of-plane angle, as we have discussed in Figure S4. In this case, the defocused distance has to be carefully adjusted to be around 0.95-1.05 μm.

To conclude, we have validated a novel non-photobleaching and non-photoblinking optical orientation sensor at nanoscale by defocused imaging of single AuNRs through a common wide field microscope. By simultaneously measuring the polarization characteristic of TPL for a large number of single AuNRs with aspect ratios of 1.2-4, we correlate their defocused images with their three-dimensional spatial orientations. Compared with previous reports, the orientation probes proposed in our paper are independent on the polarization and wavelength of the exciting source, the size can be down to ten nanometers and the distribution of the aspect ratio was examined to have negligible effects on the sensing property. Furthermore, since deciphering the images can be finished off-line at a later time and many AuNRs can be monitored parallel within one frame, video-rate acquisition of multiple AuNRs is possible and the probing is highly effective. These unique features make the orientation probes demonstrated in this paper outstanding candidate for optical imaging and sensing in material science and biology.

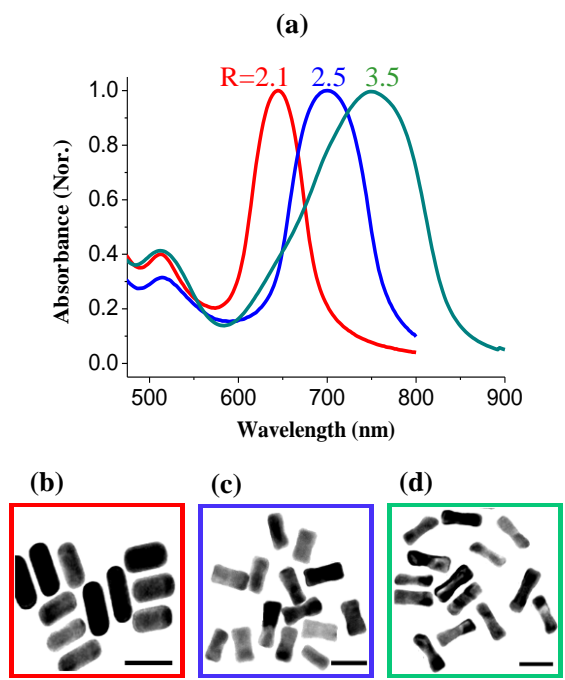

**Figure 1.** (a) Ensemble extinction spectra of AuNRs taken in aqueous solutions with average aspect ratios of 2.1 (red), 2.5 (blue) and 3.5 (green). (b-d) Corresponding representative TEM images of AuNRs. The scale bar dimension is 50 nm.

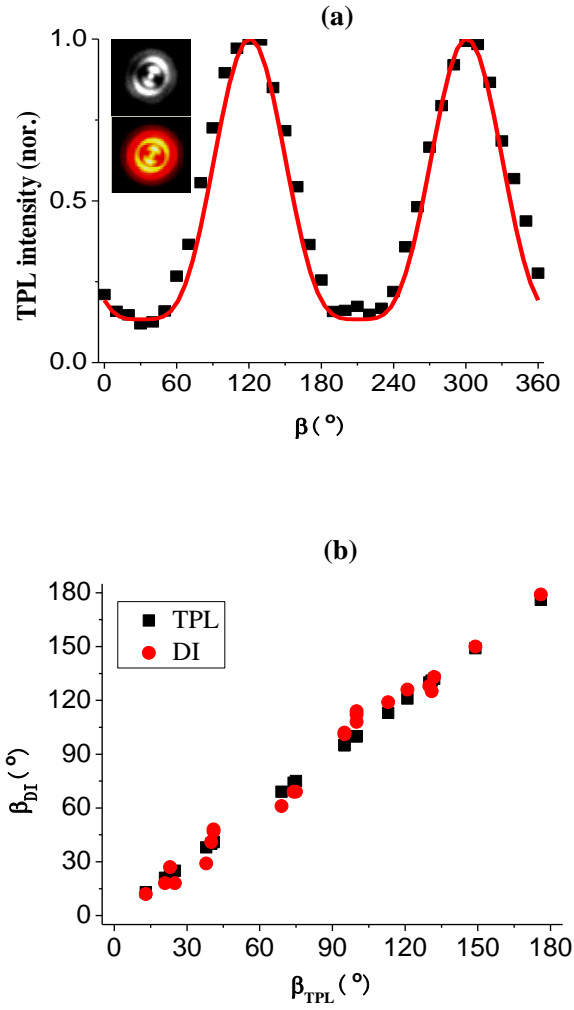

**Figure 2.** (a) Excitation polarization dependence of the TPL intensity integrated over 570-640 nm. The solid curve is fitted by $\cos^4(\theta-120.6°)$ function and β is thus determined to be 120.6°. The insets shows the measured (white and black) and simulated (red and yellow) defocused images (DI). (b) AuNR orientations obtained from the excitation polarization characteristics of TPL (black square) and defocused images (red circle).

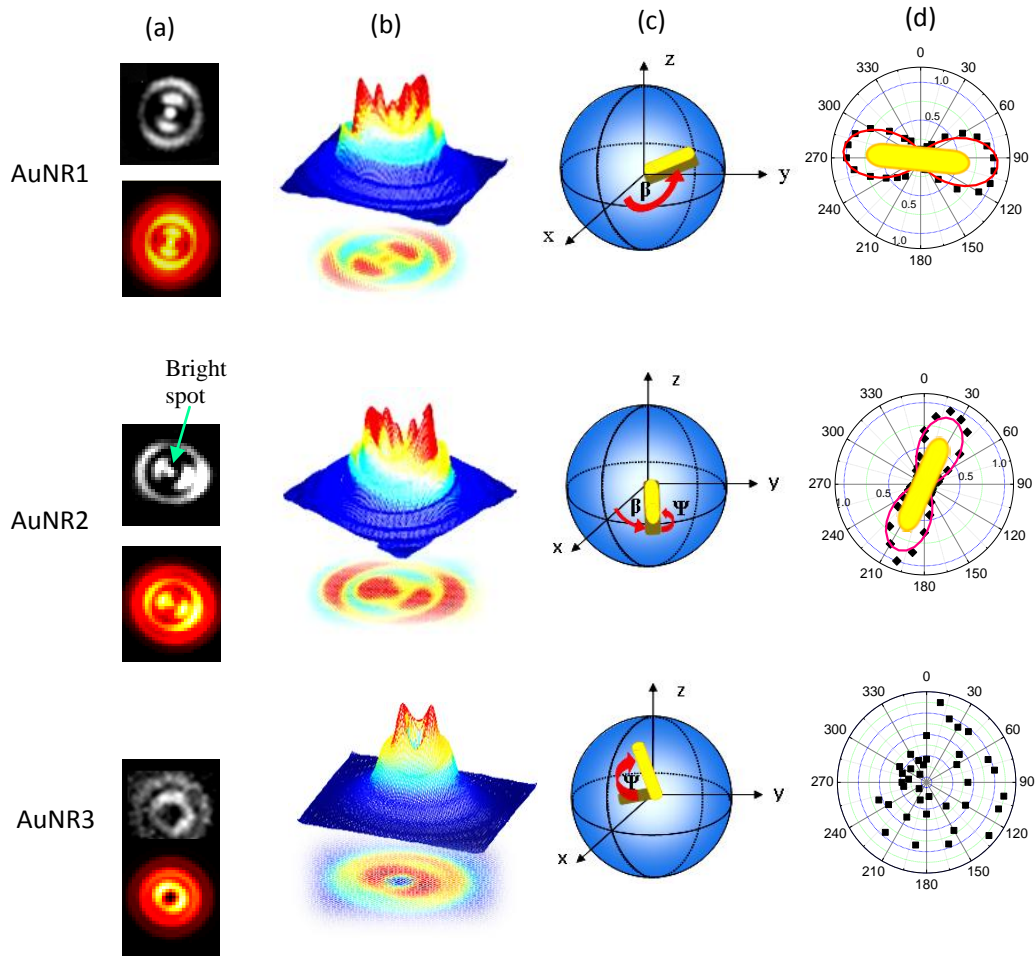

**Figure 3.** Column (a) demonstrates the measured (in white and black) and simulated (in red and yellow) defocused images for three typical AuNRs from a sample with an average aspect ratio of 3.5. The defocusing distance for AuNR1, AuNR2 and AuNR3 is 1 μm, 1.1 μm and 0.92 μm respectively. Column (b) plots the 3D intensity distribution of the measured defocused images in column (a). The 3D orientations of the single AuNRs determined from the defocused images are illustrated in column (c) as yellow rods, in which the grey rods are the corresponding in-plane projections. Column (d) shows the polar plots of the TPL intensity (black squares) versus the excitation polarization direction. The red curves are drawn according to the best-fitted $\cos^4(\theta-\beta)$ function. The orientations of the yellow rods are derived from the defocused images shown in column (a).

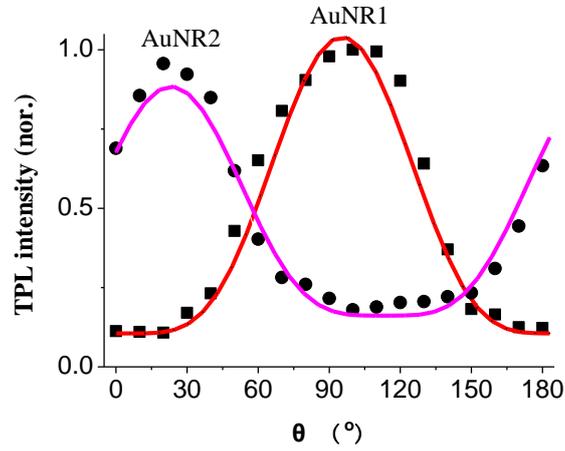

**Figure 4.** Excitation polarization dependence of the TPL intensity integrated over 570-640 nm for AuNR1 (squares) and AuNR2 (circles) demonstrated in Figure 3. The solid curves are fitted by $\cos^4(\theta-95.6°)$ (red) and $\cos^4(\theta-24.3°)$ (magenta) functions respectively. Obviously, the *DOP* of AuNR1 is larger than that of AuNR2.

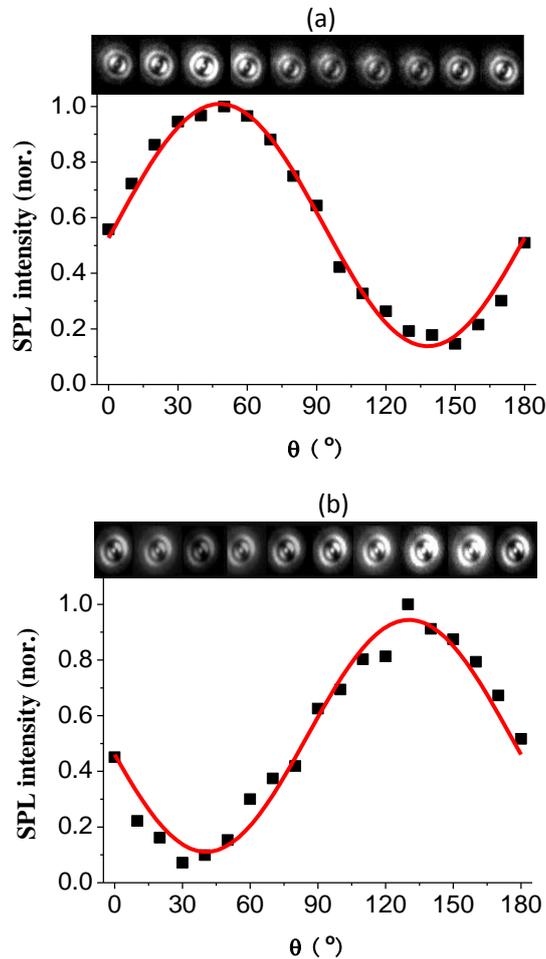

**Figure 5.** Excitation polarization dependence of SPL intensity integrated over the defocused image area and in a wavelength range 540-910 nm for two randomly selected AuNRs with different spatial orientations. The top insets are the corresponding measured defocused images. Obviously, the pattern of the defocused images remains the same although the intensity varies with the excitation polarization. (a) $\beta=41°$, $\Psi=0°$; (b) $\beta=131°$, $\Psi=0°$.

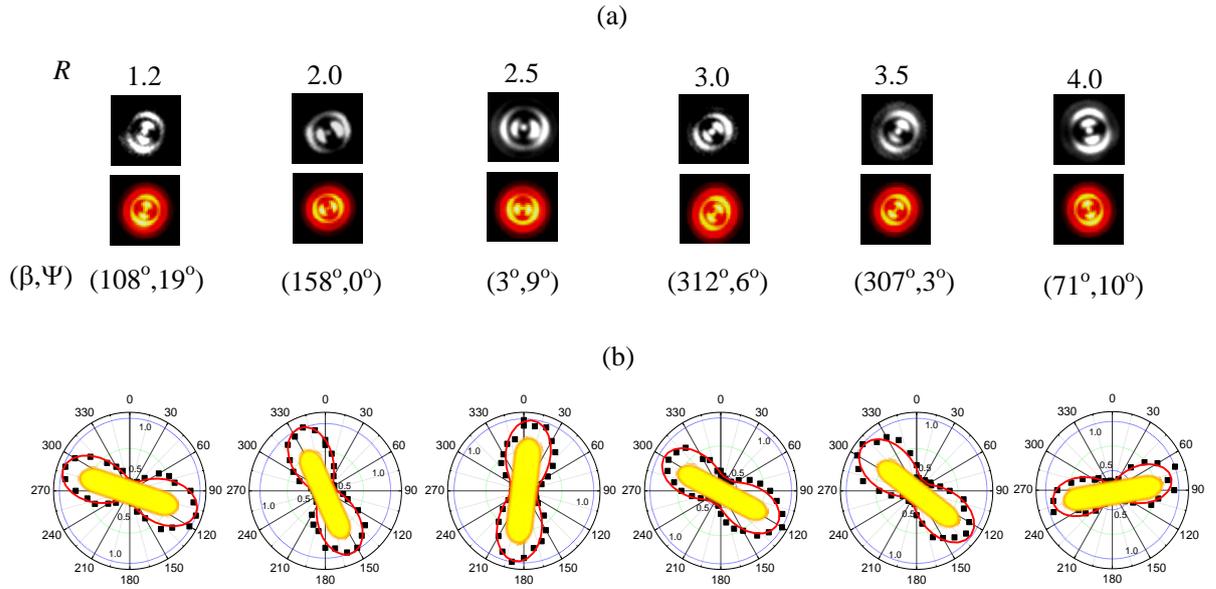

**Figure 6.** (a) Measured (top panel) and simulated (bottom panel) defocused images of single AuNRs with different aspect ratios, which are indicated at the top. (b) The polar plots of TPL intensities (black squares) versus the excitation polarization direction. The red curves are drawn according to the best-fitted $\cos^4(\theta-\beta)$ function. The orientations of the yellow rods are derived from the defocused images shown in (a).

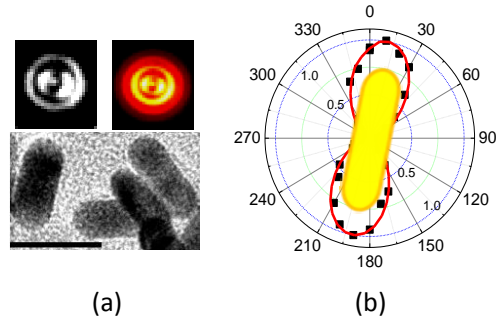

**Figure 7.** (a) A representative TEM image for the fabricated AuNRs with smaller sizes. The scale bar dimension is 20 nm. The top insets show the measured (in black and white) and simulated (in red and yellow) defocused images of a typical AuNR with the diameter and length as small as about 9 nm and 20 nm in average. $\beta$ and $\Psi$ is derived to be 10° and 14° respectively. (b) The polar plot of the TPL intensity (black squares) versus the excitation polarization direction. The red curves are drawn according to the best-fitted $\cos^4(\theta-\beta)$ function. The orientation of the yellow rod is derived from the defocused image.

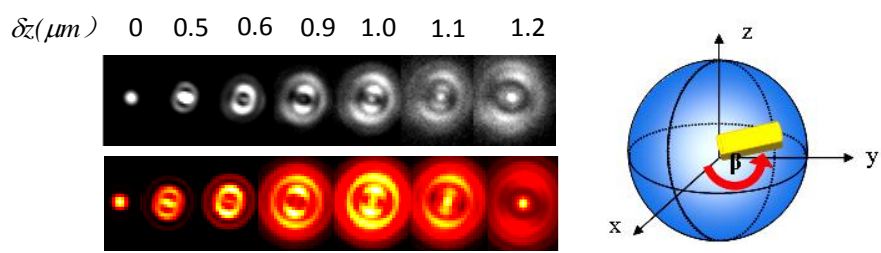

**Figure 8.** (a) Measured (top panel) and simulated (bottom panel) defocused images of the same AuNR with an aspect ratio of 2.5 at different defocusing distance $\delta z$, which is illustrated at the top. $\delta z = 0$ corresponds to the focal plane. The spatial orientation of the AuNR is schematically plotted at right.

M. Y.; Zhang, Q. Q. *Adv. Funct. Mater.* **2008**, *18*, 277-284. Hydrothermal syntheses of gold nanocrystals: from icosahedral to its truncated form.